\documentclass[12pt]{iopart}
\pdfoutput=1

\usepackage{graphics,color,rotating,epsfig}

\usepackage[english]{babel}

\usepackage{amsfonts}
\usepackage{amssymb}

\newcommand{\bbbr}{\mbox{I\hspace{-0.1cm}R}}
\newcommand{\pader}[2] {\frac {\partial #1} {\partial #2}}
\newcommand{\bx}{\mathbf{x}}
\newcommand{\brb}{\mathbf{r}}

\newcommand{\hm}{{H$_2$} }
\newcommand{\hp}{{H$_P$} }
\newcommand{\hc}{{H$_C$} }

\begin{document}

\title[A stochastic model and MC algorithm for fluctuation-induced H$_2$ formation]
{A stochastic model and Monte Carlo algorithm for fluctuation-induced H$_2$ formation on the surface of interstellar
dust grains
}

\author{K.K. Sabelfeld     }
\address{
Institute of Computational Mathematics and Mathematical Geophysics, Russian Academy of Sciences,
Lavrentiev Prosp. 6, 630090 Novosibirsk, Russia}
\ead{karl@osmf.sscc.ru}


\begin{abstract}
A stochastic algorithm for simulation of fluctuation-induced kinetics of H$_2$ formation on grain surfaces
is suggested as a generalization of the technique developed in our recent studies \cite{sab-bra-kag}   where this method was developed
to describe the annihilation of spatially separate electrons and holes in a disordered semiconductor.
The stochastic model is based on the spatially inhomogeneous, nonlinear integro-differential Smoluchowski equations with random source term.
In this paper
we derive the general system of Smoluchowski type equations for the formation of H$_2$ from two hydrogen atoms on the surface of interstellar
dust grains with physisorption and chemisorption sites. We focus in this study
 on the spatial distribution, and numerically investigate the segregation  in the case of a source with a continuous generation in time and
 randomly  distributed in space.
The stochastic particle method presented is based on a probabilistic interpretation of the underlying process as a stochastic Markov process of interacting particle system in discrete
but randomly progressed time instances. The segregation is analyzed through the correlation analysis
of the vector random field of concentrations which appears to be isotropic in space and stationary in time.

\vspace{1cm}

\end{abstract}



\maketitle

\section{Introduction}

 In the literature, there exist several theoretical models and methods to study the surface chemistry
 
that occurs on interstellar dust grains under different
astrophysical conditions. We mention here the following approaches: the conventional rate equation method
which can be implemented using the master equation, or a Monte Carlo technique which are able to
simulate the diffusive reactions on surfaces.
The kinetic Monte Carlo technique has been applied in  all branches of astrochemistry studies,
see, e.g. \cite{charnley}, \cite{cuppen}, \cite{iqbal}, \cite{lichtenberg} and seems to be the most popular
in practical calculations. The moment equation approach presents a nice analytical approach, however it
is much weakened by the closer assumptions \cite{lipshtat}.

The fluctuation impact on the kinetics in the interstellar chemistry has not been studied thoroughly before,
in spite the fact that in chemical kinetics, the segregation  has been studied by many authors, see, for instance \cite{lichtenberg}
where the segregation phenomenon has been described from a general point of view in the case of a source, random in space and continuous in time.
 Note that the Monte Carlo methods used in studies mentioned above  are of different type: they do not deal with
 the spatial fluctuations of the source, actually, they suggest a randomized solution of deterministic rate equations
 which do not govern the segregation phenomenon.

Let explain this by a simple example by considering a reaction of two types of particles,  $A$ and $B$, leading to a product $P$: \ $A + B \to P$.
The simplest kinetic approach to this reaction first considered by Smoluchowski \cite{smoluchowski}
is based on the rate equations
$$
dn_A/dt=-g\,n_A(t)n_B(t), \quad dn_B/dt=-g\,n_A(t)n_B(t), \quad n_A(0)=a_0, n_B(0)=b_0
$$
where $g$ is the reaction rate. For diffusion-controlled reactions in $\bbbr^3$,
Smoluchowski obtained $g=4\pi D r_0$, where $r_0$ is the particle radius and $D$ is the relative diffusion coefficient. This equation can be easily solved explicitly:
$
n_A(t)={a_0}/{(1+a_0gt)}$ for $a_0=b_0$, and $n_A(t)=a_0 f_0/[b_0\exp(f_0gt)-a_0]$ for $a_0<b_0$,
where $f_0=b_0-a_0$.

Such description of chemical reactions implies conditions such that the
rate at which the reactants approach each other (the diffusion rate)
is much larger than the rate at which they react chemically. Hence, the basic assumption
underlying this reaction model is a homogeneous spatial
distribution of particles during the reaction at any time instant,
i.\,e., the components $A$ and $B$ should be always perfectly mixed.
Independent of the dimension $d$, this assumption results in solutions with a long-time asymptotics $ \sim \exp(-f_0 g t)$ if $a_0\ne b_0$, and $\sim 1/gt$ if $a_0=b_0$.

 In this approach, the diffusion is treated macroscopically, ignoring density fluctuations. In the absence of nonlinear interactions such as chemical reactions, the macroscopic diffusion equations govern uniform concentration distributions. Thermal fluctuations, initial density inhomogeneities, and the randomness of reaction events lead to non-uniform concentration fields and changes the solution structure drastically, in particular,
  the time dependence of the mean solution for asymptotically long times.

Fluctuations are responsible for the spatial correlations, and in particular,
they may lead to segregation, i.\,e., the
formation of spatially separated clusters composed entirely of
particles of either type $A$ or $B$. In this paper we
of appearing segregation in terms of the input parameters.

It
was first shown by Ovchinnikov and Zeldovich \cite{zeldovich} that in the fluctuation-induced reactions,
the long-time asymptotics is $\sim t^{-3/4}$ if $a_0=b_0$. Generally, for $a_0=b_0$, the asymptotics $\sim t^{-d/4}$ is valid for any dimensionality $d\le 4$, while for $d \ge 4$, $\sim t^{-1}$. This law was obtained by several authors using different arguments (see, e.\,g., Refs.~\cite{zeldovich}, \cite{bramson},  \cite{kotomin-kuzovkov}, and \cite{sab-bra-kag}).
If the densities of the components are not equal, the asymptotics is different. For instance, if $a_0< b_0$,
$n_A \sim a_0\exp(-\sqrt{t})$ for $d=1$, $n_A \sim a_0\exp\{-t/\log(t)\}$ for $d=2$, while for $d\ge 3$, the asymptotic law again coincides with the homogeneous case:
$n_A \sim a_0\exp(-t)$ \cite{bramson}.

This difference between the homogeneous and fluctuation-limited kinetics holds regardless of the type of reaction between the particles. It applies, for instance, to the Smoluchowski coagulation equation. Coagulation, or coalescence, is a process by which two particles collide and adhere, or coagulate. There are many different mechanisms that bring two particles to each other: molecular diffusion, gravitational sedimentation, free molecule collisions, turbulent motion of the host gas, acoustic waves, density, concentration and temperature gradients, electric charges, etc.\ (see, e.\,g., Refs.~\cite{williams} and \cite{kolodko-sabelfeld}).
Let us consider the Smoluchowski equation governing the coagulation of particles of different type
having different masses $m_i$: (e.g., see \cite{sab-1998}:
\begin{eqnarray}
\label{int4}
&&\pader {n(m,t)}{t}
=
\frac{1}{2}\int\limits_0^{m_1}\ldots \int\limits_0^{m_s} R(u,m-u)n(m-u,t)n(u,t)du   \nonumber \\
&&\hspace{5cm} -n(m,t)\int\limits_0^{\infty}\ldots \int\limits_0^{\infty}
R(u,m)n(u,t)du.
\end{eqnarray}

Here, $m_i$ is the mass of the $i$-th component in a particle, and $m$ is a vector of compositions $(m_1,\ldots,m_s)$, where $s$ is the total number of components; $n(m,t)dm$ is the number
of particles having mass of component $i$ in the range $[m_i,m_i+dm_i]$ at time $t$, and $R(u,m)=R(m,u)$ is the binary coagulation coefficient.

The numerical solution of the inhomogeneous Smoluchowski equations is a highly challenging problem even for only a few particle types.
We deal in this paper with three particle types (physisorbed H (H$_P$, chemisorbed H (H$_C$), and molecular hydrogen H$_2$).

The main difficulties in the problem we solve, can be formulated as follows:
\noindent
(1) The  major difficulty arises from the inhomogeneity in space. \\
(2) The second difficulty of our problem is caused by the low and singular particle densities. \\
(3) Third, we are interested in the particle kinetics for very long times, to reach and study a quasi-stationary regime. \\
(4) Finally, we deal with stochastic source term for phisisorbed H, and we have to take the average over a reach
ensemble of source realizations. In addition, other parameters of the governing equations may
fluctuate randomly and have a large impact on the kinetics of the processes studied.

Conventional numerical methods are not applicable for handling problems of this kind, and we therefore use the Monte Carlo methods we have developed in our recent papers \cite{sab-bra-kag} and \cite{sab-lev-kir}
which were based on the approach
previously suggested in our publications
Refs.~\cite{srkl-1996}, \cite{ksw-1999}, and \cite{kolsab-2003}.

The main idea of Monte Carlo methods for solving the spatially
homogeneous Smoluchowski equation lies in the probabilistic
interpretation of the evolution of the interacting particles as a
Markov chain (see, e.\,g., Ref.~\cite{ksw-1999}). In Refs.~\cite{kolodko-sabelfeld} and \cite{kolsab-2003}, we have applied the Monte Carlo technique to the inhomogeneous Smoluchowski equation. In this paper, we  consider the general inhomogeneous case with diffusion. Note that the direct Monte Carlo simulation of the particle interactions and diffusion jumps on a grid is computationally expensive, because one has to consider a huge number of jumps per one particle interaction. In \cite{sab-bra-kag}   we suggest a new Monte Carlo method for this case, introducing "long diffusion jumps" which accelerate the simulation process significantly.
However this approach being very efficient for calculating most important statistical
characteristics of the solution, is not well adjusted when there is a need to analyze the spatial distribution,
in particular, the segregation process. Therefore, we use a detailed Monte Carlo scheme where in each random time step,
all processes including the diffusion jump are directly simulated.

\section{The inhomogeneous nonlinear Smoluchowski equation based model}

We assume that hydrogen atoms absorbed on a grain site can move to another site via tunneling or thermal diffusion.
We define $H_P$ , $H_C$ and $H_2$ as the
physisorbed H, chemisorbed H and molecular hydrogen concentrations, respectively.
We may imagine that the granular surfaces to be square lattices with four nearest neighbor sites, as
on fcc[100] plane. But we would prefer to use a continuous description, both in space and time.
This approach dramatically reduces the memory and computer time needed, and makes it possible to
simulate the H$_2$ formation of very large grain surfaces. In the case of low temperatures, when the diffusion is negligible,
we work on a grid.

Different energy barriers occur between different pairs of sites which can be sites in which H atoms are weakly bound, due to physisorption (weak Van der Waals interaction),
or strongly covalent bound, due to chemisorption.  We assume the number of physisorbed and chemisorbed sites on a
grain to be identical.
We denote by $D_P$ the diffusion coefficient for the physisorption, and by $D_C$ the chemisorption
diffusion coefficient. The recombination of spatially separate sites distributed in a grain $G$ due to tunneling from a physisorption to physisorption site is defined by the rate
$\alpha_{PP}(\brb)$,
from a chemisorption to physisorption site it is defined by $\alpha_{CP}(\brb)$. Analogously,
 the tunneling from a chemisorption to chemisorption site is defined by the rate $\alpha_{CC}(\brb)$,
from a physisorption site to  chemisorption site is defined by the rate $\alpha_{PC}(\brb)$. These tunneling coefficients have the form $\alpha_{ij}(\brb)=\alpha^0_{ij}\exp(-|x|/a_{ij})$
where $|x|$ is the distance between the two interacting sites. Here $a_{ij}$ is the characteristic distance of tunneling,
and  $\alpha^0_{ij}$ (i,j=P,C) are the frequencies of the relevant events which are described in details in \cite{iqbal}, see also \cite{cazaux}.

The spatially varying desorption rates are designated
by $W_{H_P}$ and $W_{H_C}$. The source $F(\brb)$ is the accretion rate.
In this study we assume that $F(\brb)$ is a spatial random field, and in simplest case
we use the Poissonian distribution.


Altogether, the concentrations $H_P(\brb,t)$ and $H_P(\brb,t)$
are governed by the following system of two coupled nonlinear Smoluchowski equation

\begin{eqnarray}
\label{eq1}
&&
\hspace{-2cm}
\frac{\partial H_P(\brb; t)}{\partial t}
= D_{P}(\brb) \Delta H_P(\brb; t) -\alpha_{PC}H_P- H_P(\brb; t) \int \alpha_{PP}(|\bx|)H_p(\brb + \bx; t)d\bx \nonumber \\
&&
\hspace{-2cm}
+\alpha_{CP}H_C - H_P(\brb; t) \int \alpha_{CP}(|\bx|)H_C(\brb + \bx; t)d\bx
   - W_{H_P}H_P +F(\brb)~,
\end{eqnarray}
  \begin{eqnarray}
\label{eq2}
&&
\hspace{-2cm}
\frac{\partial H_C(\brb; t)}{\partial t}
= D_{C}(\brb) \Delta H_C(\brb; t) -\alpha_{CP}H_C- H_C(\brb; t) \int \alpha_{CC}(|\bx|)H_C(\brb + \bx; t)d\bx \nonumber \\
&&
+\alpha_{PC}H_P - H_C(\brb; t) \int \alpha_{PC}(|\bx|)H_P(\brb + \bx; t)d\bx
   - W_{H_C}H_C~.
\end{eqnarray}

Without loss of generality, we assume that the total initial concentrations of $H_P$ and of $H_C$ are equal.
At the initial time $t = 0$, the concentrations $H_P$ and $H_C$    are zero.

Since the source $F$ is a random field, so are the concentrations $H_P$ and $H_C$. Therefore,
the experimentally measured is the mean concentration of the molecules $H_2$.
The kinetics of $H_2$ concentration reads
\begin{eqnarray}
\label{eqH2}
&&
\hspace{-2.5cm}
\frac{\partial H_2(\brb; t)}{\partial t}=H_P(\brb; t) \int \alpha_{PP}(|\bx|)H_p(\brb + \bx; t)d\bx+
H_P(\brb; t) \int \alpha_{PC}(|\bx|)H_C(\brb + \bx; t)d\bx \nonumber \\
&&
\hspace{-2.0cm}
+ H_{C}(\brb; t) \int \alpha_{CP}(|\bx|)H_P(\brb + \bx; t)d\bx +
H_C(\brb; t) \int \alpha_{CC}(|\bx|)H_C(\brb + \bx; t)d\bx~.
\end{eqnarray}

We are interested both in kinetics and the quasi-steady state which is reached
in which the total  mean surface population of H atoms fluctuates around a constant value.
After the steady-state has been reached
the H$_2$ formation efficiency is then defined by the ratio of the means:
\begin{equation}\label{meanh2}
\eta=\frac{1}{\langle \int_G F(\brb) d\brb \rangle } \,\, \langle  \int 2 H_2(\brb)\, d\brb \rangle~.
\end{equation}
where the angle brackets stand for the ensemble average generated by the random source F.

\section{Monte Carlo Algorithms}
\sloppypar
The equations (\ref{eq1})--(\ref{eq2}) have the structure of inhomogeneous
Smoluchowski coagulation equations, as mentioned in the introduction.
The Smoluchowski equations can be interpreted probabilistically as an equation generated by Markov chains describing the evolution of
pairwise  interacting particle system. In Ref.~\cite{kolodko-sabelfeld}), we developed a Monte Carlo algorithm for inhomogeneous Smoluchowski equation,
which we adapt in \cite{sab-bra-kag}. Here it will be used to solve Eqs.~(\ref{eq1}-\ref{eq2})
for the  two-dimensional case  $d=2$, with the focus on the segregation problem.

\subsection{Recombinations by tunneling, in the absence of diffusion}
\label{subsec41}

For simplicity, let us first consider the algorithm for the case when there is no
diffusion. The direct simulation assumes that the process of
interactions  is pairwise and Markovian, i.\,e., having made a
time step, the next time step is simulated independently. The
interacting pair is sampled according to the kernel of the equation
$B=B_0\exp(-|\bx|/a)$ as described below.

Let us assume our phase space is continuous. The process is
simulated in a square box with size $L\times L$ and periodic boundary
conditions.

In the first step, we sample between the following possible events: (\emph{i}) formation of H$_2$ by tunneling
from H$_P$ to H$_P$,  (\emph{ii}) formation of H$_2$ by tunneling from H$_P$ to H$_C$, (\emph{iii})
formation of H$_2$ by tunneling from H$_C$ to H$_C$, and (\emph{iv})
formation of H$_2$ by tunneling from H$_C$ to H$_P$. The processes are described by the relevant
probabilities $\alpha_{ij}(\brb)=\alpha^0_{ij}\exp(-|x|/a_{ij})$.
Thus, the simulation algorithm in the absence of diffusion can be described as follows.

\begin{enumerate}
	\item Put $t=0$,
	and sample $n=n_0$ atoms \hp and $p=n_0$ atoms \hc at random, independently and uniformly distributed in the box.
	\item Sample one of the possible events: (\emph{i}) \hp+\hp, (\emph{ii}) \hp+\hc, (\emph{iii}) \hc+\hc and (\emph{iv}) \hc+\hp.
To do this, first calculate the majorant frequencies for the four events:
	\begin{eqnarray}\label{a1}
	\lambda_1&=\frac{n(n-1)}{2} \alpha^0_{PP} \exp{\{-\frac{r_{nn,min} }{a_{PP}} \}},\\
	\lambda_2&=np \alpha^0_{PC} \exp{\{-\frac{r_{np,min} }{a_{PC}} \}},\\
	\lambda_3&=np \alpha^0_{CP} \exp{\{-\frac{r_{np,min} }{a_{CP}} \}},\\
	\lambda_4&=\frac{p(p-1)}{2} \alpha^0_{CC} \exp{\{-\frac{r_{pp,min} }{a_{CC}} \}},
\end{eqnarray}
	where $r_{np,\min}$ is the minimal of all possible distances between $n$ \hp atoms and $p$
	\hc atoms in the box, and analogously for other cases.
	From these frequencies, calculate the probabilities $p_1, p_2, p_3, p_4$ of the events (\emph{i}), (\emph{ii}), (\emph{iii}) and (\emph{iv}), respectively:
	$p_1=\lambda_1/\lambda$, $p_2=\lambda_2/\lambda$, $p_3=\lambda_3/\lambda$
	and $p_4=1-p_1-p_2-p_3$, where $\lambda=\lambda_1+\lambda_2+\lambda_3+\lambda_4$.
	\item From the probabilities $p_1, p_2, p_3, p_4$, sample the event $k=i, ii, iii, iv$, calculate the time increment
	as $\Delta t=-\log(rand)/\lambda$, and calculate $t:=t+\Delta t$.
	\item For the sampled event $k (=i,ii,iii,iv)$, choose uniformly the relevant interacting pair, and check if the
	interaction takes place.  For instance, if $k=ii$, i.\,e., the sampled event happens to be recombination \hp+\hc, calculate $\bar P_{np}=\exp{ \{\frac{-r+r_{np,min} }{a}\}}$. If $rand< \bar P_{np}$, then the event occurs. Hence, recalculate $n:=n-1$ and $p:=p-1$, and go back to step 2.
	Otherwise, nothing happens, the probabilities $p_1, p_2, p_3,p_4$ remain the same,	and so return to step 3. Here $r$ is the distance between the sampled interacting pair. For the case $k=i$,  put $n:=n-2$.
	If the concentrations $n$ and $p$ are calculated at some prescribed time instances $t_m$, $m=1,\ldots M$,
	just score the values $n(t_m), p(t_m)$,  $m=1,\ldots M$.
	To calculate the concentration of \hm, say, at a time  $t$ from the interval $t\in [t_1,t_2]$,	 count $M$, the number of all
 recombinations which have occurred during this time interval, and take the approximation $I(t)\approx M$.
	\item To carry out the average, run the steps 1--4 independently, say, $\nu$ times, with $\nu$ being a sufficiently large number, and take the arithmetic mean.
\end{enumerate}

Note that there is no need to recalculate the value $r_{\min}$ after each  tunneling: this should be done only if the pair with $r_{\min}$ reacts.

\subsection{Recombination in the presence of diffusion}

In the general case when the atoms \hp and \hc not only recombine, but also diffuse, the algorithm becomes more sophisticated.
 Usually, we would consider diffusion to occur by microscopic random jumps according to the law $dl=\omega\,\sqrt{D\,dt}$, where $dl$ is the length of the random jump, $\omega$ is a random (isotropic) direction, and $D$ is the diffusion coefficient. However, since the time between individual recombination events may be very large compared to $dt$, a huge number of diffusion jumps would be required to simulate the recombination dynamics.

To accelerate this algorithm, we have suggested in \cite{sab-bra-kag}
  a Random Walk on Spheres based algorithm (more details about the Random Walk on Spheres method can be found in Ref.~\cite{sabelfeld-book}). The idea behind this method is simple. Around each diffusing atom, we construct a disk of maximal radius which does not contain any atom
  \hp or \hc. We then simulate the random exit times $\tau_k$ of the atom from these disks. We chose the atom which has a minimal exit time, and let this atom jump out of the disk, so that the new random position of the atom is uniformly distributed on the boundary of this disk.
The distribution of the exit time is known (e.g., see \cite{sab-bra-kag}, and we are thus able to simulate the random time $dt$ according to this distribution, giving us the time $t:=t+dt$.

This method has shown high efficiency for calculation of the intensity, as reported in \cite{sab-bra-kag}.
For our purpose  however this approach cannot be directly applied: for
viewing at the aggregation process where clusters of \hp atoms are separated by the clusters of \hc atoms,
we need to simulate the process step by step, with sufficiently small time intervals.
Therefore, we have introduced a mesh, and the atoms were diffusing over the mesh with the frequency
$\lambda_5=D n$ for \hp, and $\lambda_6=D p$ for \hc.

The general code described  above remains the
same. We have only to sample an additional event that the atom \hp
makes a jump with frequency $\lambda_5$, and the atom \hc makes  a jump with frequency $\lambda_6$. Then, taking
$\lambda_5$ and $\lambda_6$ as described above, put $\lambda=\sum_{i=1}^6\lambda_i$, and calculate
the probabilities  as $p_i=\lambda_i/\lambda$, $i=1,\ldots 6$.

Analogously the resorption and exchange of \hc and \hp positions are simulated, using the rates $W_{H_P}$, $W_{H_C}$, and
$\alpha_{PC}, \alpha_{CP}$, respectively, which introduces four more events. Finally, the event of appearance
a new \hp atom is simulated according to the accretion rate, or flux F, of H atoms onto
the surface of a dust grain. Thus in total we have 11 events to be sampled at each time step.

\section{Simulation results}

As mentioned in the introduction, inhomogeneous fluctuations lead to the formation of clusters.
The clustering slows down the reaction considerably, because only particles near the
boundary of the clusters are likely to react, while particles inside
the cluster have to diffuse to the boundary before they have a chance
to react with a particle of the other type. In other words,
fluctuations induce the formation of a mosaic of continuously growing
domains which contain only one of the two components, \hp or \hc.

Our simulations are adapted to take place on a spatial and temporal scale comparable to that for the
surface of interstellar dust grains, given in  \cite{iqbal} and \cite{cazaux}.
In Figure 1 we show 6 samples of random concentration fields of \hp and \hc
when the system reaches a stochastically stationary regime. The segregation phenomen is clearly seen:
large clusters of \hp and \hc are negatively correlated, and should be taken into account when
evaluating the concentration of the molecular hydrogen from the equation (\ref{eqH2}). Without this correction,
the model would much overestimate the efficiency $\eta$ given by (\ref{meanh2}).
The evaluation of $\eta$ in the framework of this model will be done in the forthcoming paper.

\begin{figure}[!ht]
\hspace{-2cm}
\begin{minipage}[c]{.75\linewidth}
\includegraphics[height=6.8cm,width=22cm]{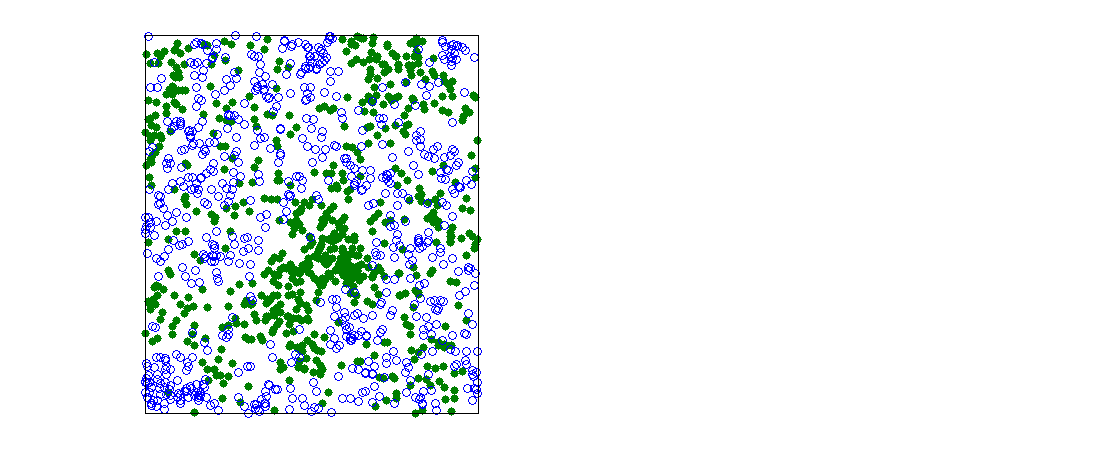}
\end{minipage}
\hspace{-3cm}
\begin{minipage}[c]{.75\linewidth}
\includegraphics[height=6.8cm, width=22cm]{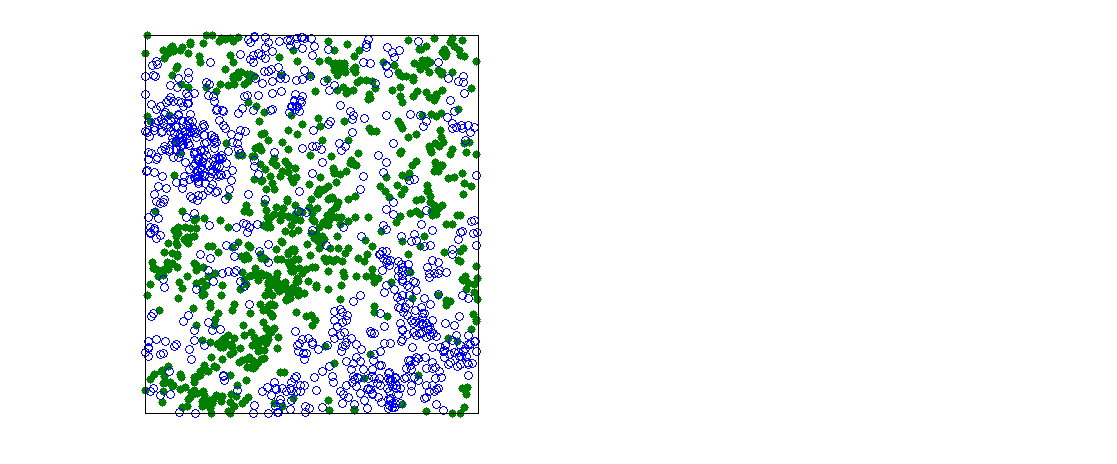}
\end{minipage}

\hspace{-2cm}
\begin{minipage}[c]{.75\linewidth}
\includegraphics[height=6.8cm,width=22cm]{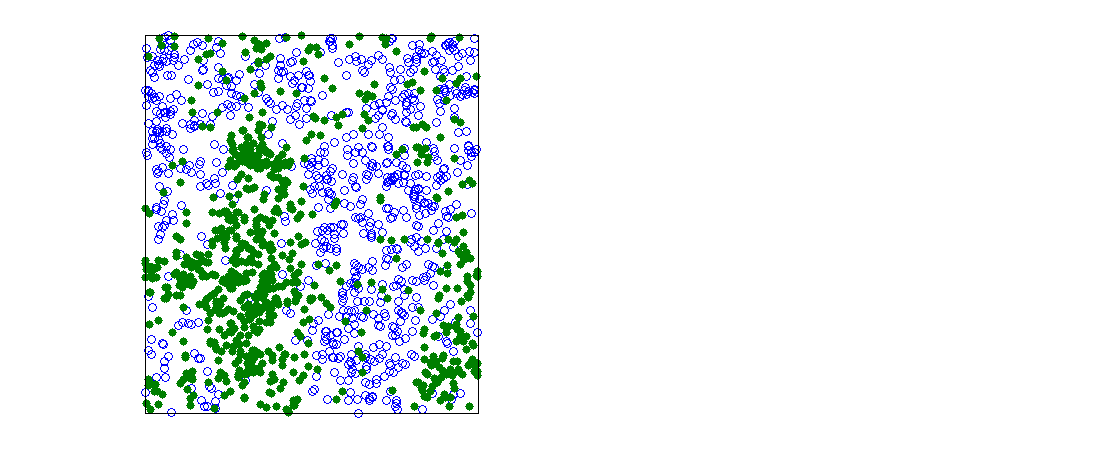}
\end{minipage}
\hspace{-3cm}
\begin{minipage}[c]{.75\linewidth}
\includegraphics[height=6.8cm, width=22cm]{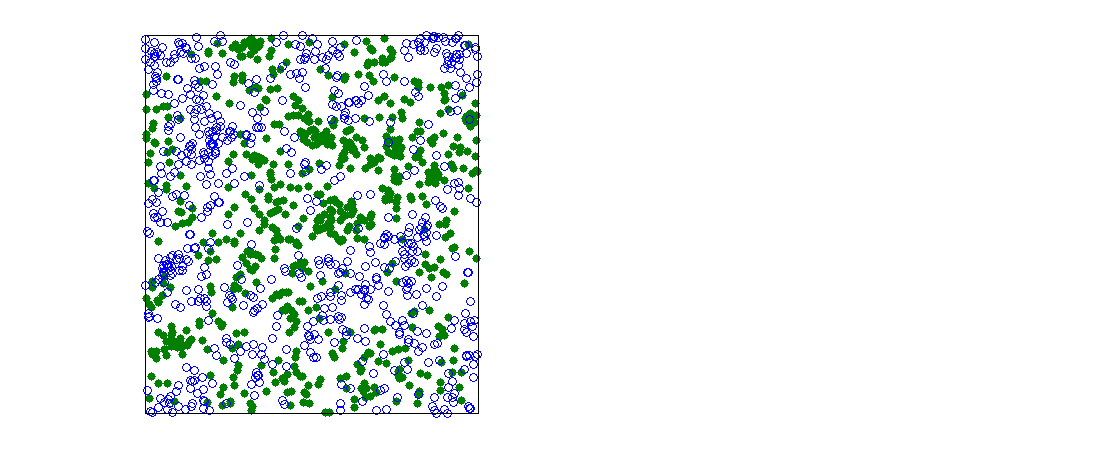}
\end{minipage}

\hspace{-2cm}
\begin{minipage}[c]{.75\linewidth}
\includegraphics[height=6.8cm,width=22cm]{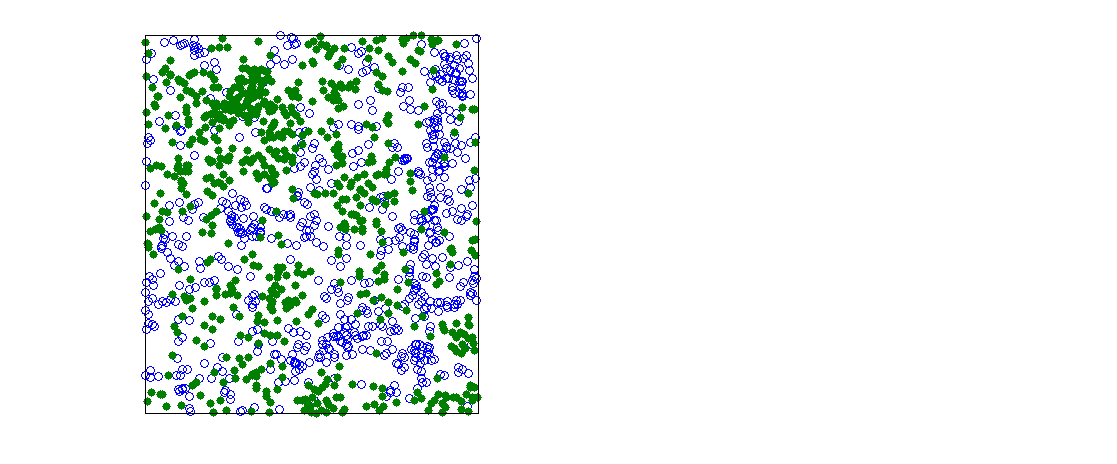}
\end{minipage}
\hspace{-3cm}
\begin{minipage}[c]{.65\linewidth}
\includegraphics[height=6.8cm, width=22cm]{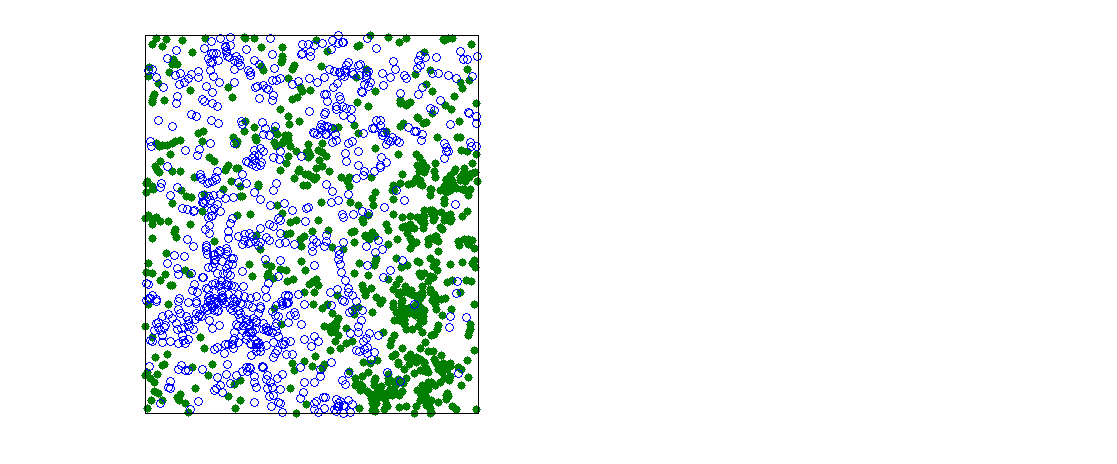}
\end{minipage}
\caption{\small{
Six  samples of \hp and \hc atoms during the stochastically stationary regime on a grain of radius 0.1 $\mu$m. It is seen, that a
segregation is formed with a characteristic distance between large clusters of \hp and \hc atoms.
 }}
\label{fig2}
\end{figure}

\clearpage

\section{Summary and conclusion}
Stochastic model and Monte Carlo simulation algorithm are constructed for solving a nonlinear system of inhomogeneous 2D Smuluchowski equations with random source term
for simulation of H$_2$ formation on grain surfaces.
The general system of inhomogeneous Smoluchowski type equations is used to govern the formation of H$_2$ from two hydrogen atoms on the surface of interstellar
dust grains with physisorption and chemisorption sites. Both tunneling and diffusion mechanisms are taken into account.
We focus in this study
 on the spatial distribution, and numerically investigate the segregation  in the case of a source with a continuous generation in time and
 randomly  distributed in space.
The stochastic particle method presented is based on a probabilistic interpretation of the underlying process as a stochastic Markov process of interacting particle system in discrete
but randomly progressed time instances. The segregation is analyzed through the correlation analysis
of the vector random field of concentrations which appears to be isotropic in space and stationary in time.
Note that the suggested model can be easily extended to more general situations, in particular,
additional capture centers may be introduced in the system where the \hp and \hc may recombine without
giving a contribution to the molecular hydrogen. This model will be presented in the forthcoming paper.

\ack{The author kindly acknowledges the help of A. Kireeva in computer simulations, 
and support of the Russian Science Foundation under Grant  14-11-00083.}

\end{document}